\documentclass[a4paper,preprint,preprintnumbers,amsmath,amssymb,superscriptaddress]{revtex4}

 \usepackage[T1]{fontenc}
 \usepackage{hyperref}
 \usepackage{array}
 \usepackage{amsmath}
\usepackage{tabularx}
 \usepackage{graphicx}
\usepackage{multirow}
 \usepackage{SIunits}
 \usepackage{ulem}
 \usepackage{braket}
 \usepackage{color}
 \usepackage[draft]{fixme}
\usepackage{amssymb}
\usepackage{dcolumn}
\usepackage{bm}
\usepackage{footmisc}
\usepackage{hyperref}

\newcolumntype{P}[1]{>{\centering\arraybackslash}p{#1}}

\begin{document}

\begin{abstract}

We demonstrate Anderson localisation of visible light on a chip and report quality factors exceeding highly engineered two-dimensional cavities. Our results reverse the trend, observed so far, of the quality of disorder-induced light confinement being orders of magnitude lower than engineered devices. Furthermore, by implementing a sensitive imaging technique, we directly visualise the localised modes, determine their position on the device and measure their spatial extension. Our findings prove the potential of disorder-induced localised light for scalable, room temperature, optical devices, operating in the visible range of wavelengths. 

\end{abstract}

\title{Anderson localisation of visible light on a nanophotonic chip} 

\today
\author{Tom Crane}
\affiliation{Department of Physics and Astronomy, University of Southampton, Southampton SO17 1BJ, United Kingdom\\ Group website: \href{http://www.quantum.soton.ac.uk}{www.quantum.soton.ac.uk}}
\author{Oliver J. Trojak}
\affiliation{Department of Physics and Astronomy, University of Southampton, Southampton SO17 1BJ, United Kingdom\\ Group website: \href{http://www.quantum.soton.ac.uk}{www.quantum.soton.ac.uk}}
\author{Luca Sapienza}\email{l.sapienza@soton.ac.uk}
\affiliation{Department of Physics and Astronomy, University of Southampton, Southampton SO17 1BJ, United Kingdom\\ Group website: \href{http://www.quantum.soton.ac.uk}{www.quantum.soton.ac.uk}}

\maketitle

Technological advances allow us to control light at the nanoscale and to strongly enhance the light-matter interaction in highly engineered devices \cite{cavities, Noda}. Controlling the propagation of visible light on a chip is of tremendous interest in research areas such as energy harvesting, imaging, sensing and biology. However, compared to state-of-the-art two-dimensional optical cavities operating at longer wavelengths, the quality factor of on-chip visible light confinement is several orders of magnitude lower. 
An alternative to highly engineered cavities for the confinement of waves is represented by disordered systems: whenever imperfections or defects are present, a wave propagating through a medium can undergo multiple scatterings. If the amount of scattering is strong enough, the wave can be trapped and its propagation can even come to a complete stop. A phase transition from ballistic propagation to localisation can thus occur \cite{Sheng} and we can enter the realm of Anderson localisation \cite{Anderson}. Such a regime can be sustained by any kind of wave and has been demonstrated for electrons, sound waves, Bose-Einstein condensates and optical waves \cite{Ad}. So far, photonic devices based on Anderson localisation have been mostly developed in the near-infrared  range of wavelengths, in particular in the telecommunication band in silicon \cite{Topolancik, Shayan, Chee}, where optical losses are the lowest, and in gallium arsenide samples containing light emitters \cite{me, Wiersma, Chee2}. The use of multiple scattering of visible light has exciting prospects in a range of applications, following theoretical predictions in energy harvesting \cite{energy} and proof-of-principle results in imaging \cite{imaging}. In this work, we demonstrate that disorder-induced high-quality confinement of visible light can be achieved at room temperature, on a silicon nitride chip.

Silicon nitride is an advantageous material for photonic applications given its compatibility with existing semiconductor fabrication technology and its transparency to a wide range of wavelengths, including the visible. Examples of its applications are integrated non-linear optics \cite{SiN} and opto-mechanical \cite{Marcelo} devices. Depending on the growth technique used, silicon nitride can host defect centres with various levels of brightness that can exhibit a broad luminescence, covering wavelengths that typically range from about 500 to 800\,nm \cite{intrinsic_PL}. Taking advantage of this, we have fabricated suspended silicon nitride photonic crystal waveguides on a silicon substrate and used the material's light emission to reveal the formation of localised modes due to the multiple scattering of light on imperfections. Using the intrinsic photo-luminescence, we image and spectrally characterise disorder-induced localised modes: to our knowledge, this is the first demonstration of Anderson localisation of visible light and of its direct imaging on a chip. Remarkably, the quality factor of the disorder-induced light confinement exceeds values reported for highly engineered two-dimensional photonic crystal cavities \cite{Benson}, making it a resource for the realisation of highly-sensitive devices.

We use a room-temperature confocal micro-photoluminescence set-up (see Fig.\,1a) where two light sources, a blue (440\,nm central wavelength) light emitting diode (LED) and a 473\,nm continuous-wave laser, are focused to illumination areas with diameters of about 50\,$\mu$m and 1.5\,$\mu$m respectively, on a sample placed on an $xy$-positioner. The LED is used to excite the photo-luminescence that is then imaged with an Electron Multiplying Charge Coupled Device (EMCCD). The laser is used to locally excite the waveguides, whose emission is then spectrally characterised via a grating spectrometer equipped with a silicon CCD. An example of the broad-range photo-luminescence spectrum measured from the bulk silicon nitride sample under investigation is shown in Fig.\,1b. By means of finite-difference time-domain simulations, we design suspended photonic crystal waveguides confining light at 650\,nm (see Fig.\,1c), the peak wavelength of the silicon nitride photo-luminescence.

When disorder is introduced in a photonic crystal waveguide, for instance by displacing the position and/or modifying the shape of the air holes, light propagation can be strongly affected and localisation can occur, giving rise to sharp spectral resonances \cite{Savona}. To this end, we fabricate waveguides with different amounts of disorder, introduced by randomly displacing the air holes in the photonic crystal with respect to the perfectly periodic structure, following a Gaussian distribution with a varying standard deviation, expressed as percentage of the photonic crystal lattice parameter. The waveguides are then illuminated with the blue LED and the photo-luminescence images are collected by the EMCCD, as shown in Fig.\,2. Thanks to the spontaneous luminescence of the silicon nitride material that acts as a homogenous internal light source, we have direct access to the confined optical modes that appear as bright spots on our imaging camera. We can thus visualise the Anderson-localised modes and map their spatial extension, a feature only made possible by the unique properties of the nanophotonic device under study. When imaging waveguides where no intentional disorder is introduced, we observe a bright photo-luminescence signal from the central waveguide channel, proving the successful guiding of light along the line defect in the photonic crystal structure. When disorder is deliberately introduced in the system, localised areas of higher intensity are visible along the waveguide, which are the signature of light confinement within optical cavities in the Anderson-localised regime (see Fig.\,2a, left panels). From the linecuts of the photoluminescence images, we can extract the far-field spatial extension of the modes, as shown in the right panels of Fig.\,2a. By analysing the statistics as a function of disorder (see Fig.\,2b), we observe that the extensions of the optical resonances, each characterised by a specific spectral signature, all lie below 1.2\,$\mu$m. These values are significantly shorter than the 100\,$\mu$m-length of the fabricated waveguides, thus proving localisation.

By locally exciting the sample with the laser source, we can probe specific positions along the waveguides and spectrally characterise the emission of the localised modes. Sharp resonances, the signature of light confinement, clearly appear superimposed on the broad silicon nitride photo-luminescence. When scanning the laser along the waveguide, multiple resonances appear at different wavelengths, as expected from confinement due to a multiple-scattering process. We typically observe 3 to 6 peaks in each spectrum: each 100\,$\mu$m-long waveguide can thus host as many as 300 disorder-induced cavity modes. We measure linewidths $\Delta\lambda$ of the resonances as low as 0.1\,nm, reaching quality factors $Q = \lambda / \Delta\lambda$ (where $\lambda$ is the central wavelength of the spectral peak) of the confined light as high as 7600 (see Fig.\,3a). This value largely surpasses the ones reported for engineered two-dimensional photonic crystal cavities in silicon nitride where typical experimental quality factors range between a few hundred \cite {Q} and one thousand \cite{Q1, Q2}, and even exceeds the previous record value of a few thousand reported for an engineered photonic crystal heterostructure cavity \cite{Benson}. To our knowledge, this represents the highest quality factor reported for two-dimensional photonic crystal cavities in the visible range of wavelengths. If we consider the ratio $Q_{A}/Q_{2D}$ between the highest quality factors reported for Anderson-localised and engineered two-dimensional cavities, at telecom wavelengths this ratio has the value of 0.02 \cite{Topolancik, 9Mill}, in the near-infrared 0.3 \cite{Hennessy, me}, while in the visible we reach 2.2 \cite{Benson}, thus reversing the trend reported so far. Remarkably, we show that disorder-induced localisation of light provides quality factors exceeding those of engineered cavities.

As opposed to engineered photonic crystal cavities, in our devices the existence of optical cavities relies on light being trapped by scattering on imperfections: a distribution of cavity wavelengths and quality factors is therefore expected and an example of its statistics is shown in Fig.\,3b. We observe a clear dependence of the quality factor of the light confinement on the degree of disorder: we measure quality factors between 70 and 7600, that decrease when the level of imperfection in the system is increased. This can be attributed to the fact that, if more defects are present, the probability for light to be scattered out of the cavity increases, thus reducing the light-confinement quality factor. As shown in Fig.\,3b, the intrinsic fabrication imperfections (0\% disorder) are enough to provide high-quality light confinement and no further engineering of the disorder is required in order to achieve the optimal operation condition for the fabricated devices.

In conclusion, we have demonstrated Anderson localisation of visible light on a silicon nitride chip and, by means of photo-luminescence imaging, we have visualised the confined optical modes and extracted their far-field spatial extension. The spectral characterisation of the localised modes has revealed confinement with high quality factors, exceeding values reported for engineered two-dimensional photonic crystal cavities. The achievement of high quality factors in the confinement of visible light can find applications in energy harvesting, imaging, sensing and fundamental research in light-matter interaction, like cavity quantum electrodynamics experiments \cite{QED} with emitters in the visible range, such as colloidal quantum dots \cite{colloidal} or defect centres in diamond \cite{diamond}. The latter are particularly attractive for quantum technology applications, thanks to their room temperature operation and spin-coherence properties \cite{diamonds}. Since diamond is a very difficult material to process, hybrid silicon-diamond structures are appealing to control light propagation and the spontaneous emission rate of defect centres \cite{hybrid_diamond}. Furthemore, given that many spatially-extended photonic modes spontaneously appear along the waveguides, their mutual overlap could be used to realise quantum information networks for the propagation of quantum light \cite{networks}, based on coupled Anderson-localised photonic modes, known as necklace states \cite{necklace}. 

\section*{Acknowledgments}

We would like to thank Kartik Srinivasan for his continuous support and for the critical reading of the manuscript, Marcelo Davan\c{c}o for useful discussions and Henry Nelson for his early contribution to this work.

\begin{figure}[htbp!]
\centering
\includegraphics[width=0.9\linewidth]{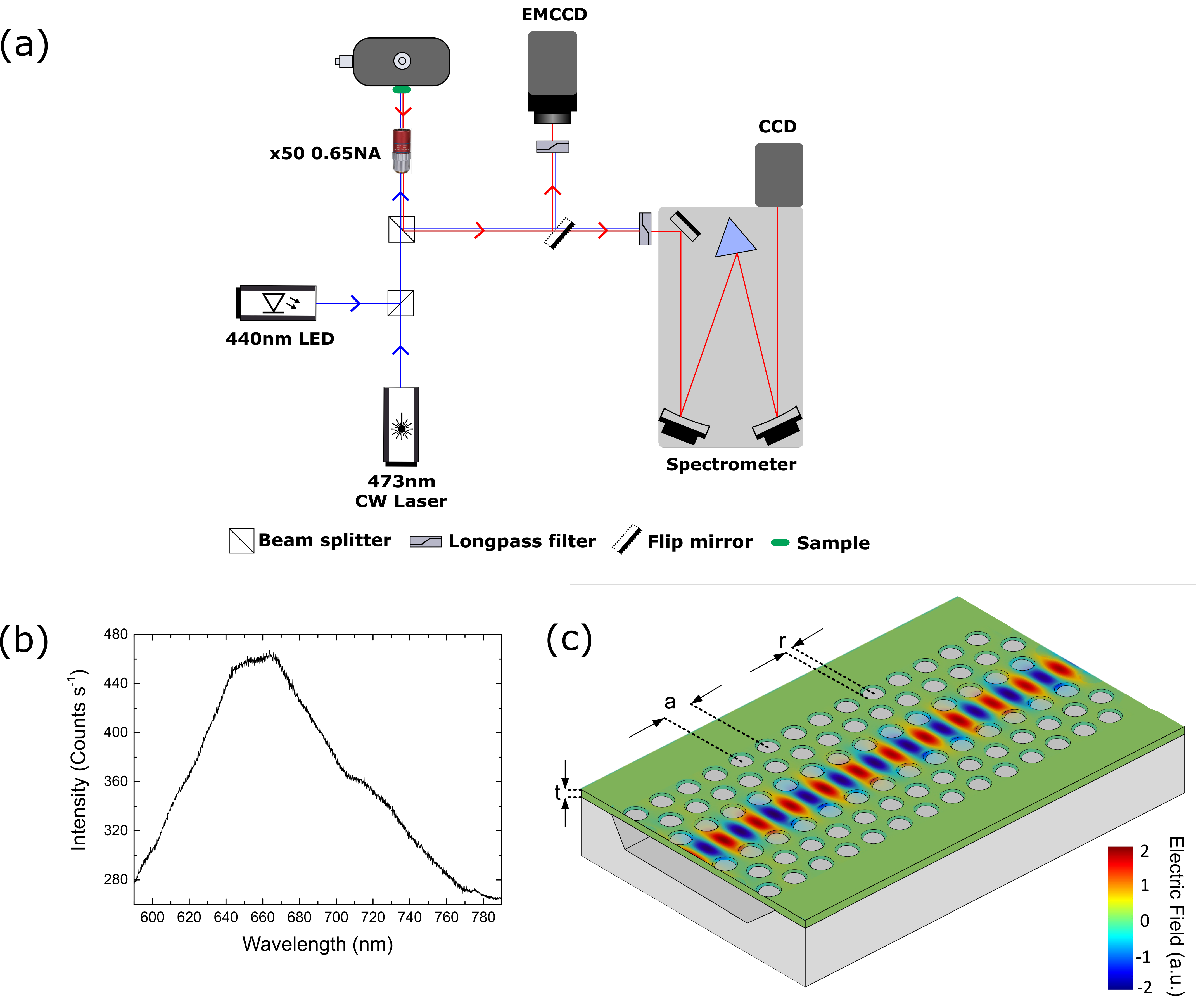}
\caption{(a) Schematic of the confocal micro-photoluminescence set-up (not to scale), comprising a light emitting diode (LED) emitting around 440\,nm and a continuous wave (CW) laser emitting at 473\,nm as excitation sources, focused by a microscope objective (with numerical aperture NA = 0.65) onto a sample placed on an $xy$-translation stage. The detection is carried out by an Electron Multiplying Charge Coupled Device (EMCCD) for photoluminescence imaging and by a CCD at the exit port of a reflection grating spectrometer for spectral characterisation. (b) Photoluminescence spectrum collected from an unpatterned area of the silicon nitride wafer, at room temperature, under laser excitation with a power density of 28\,kW/cm$^2$. (c) Schematic of the photonic crystal structure (not to scale) illustrating the lattice constant $a$, hole radius $r$ and thickness $t$ of the suspended membrane (that in the fabricated devices are 310 nm, 110 nm and 250 nm, respectively). A finite-difference time-domain simulation, depicting the electric field emitted by an embedded dipole and the resulting guided photonic mode, is superimposed on the structure.}
\end{figure}

\begin{figure}[htbp]
\centering
\includegraphics[width=0.9\linewidth]{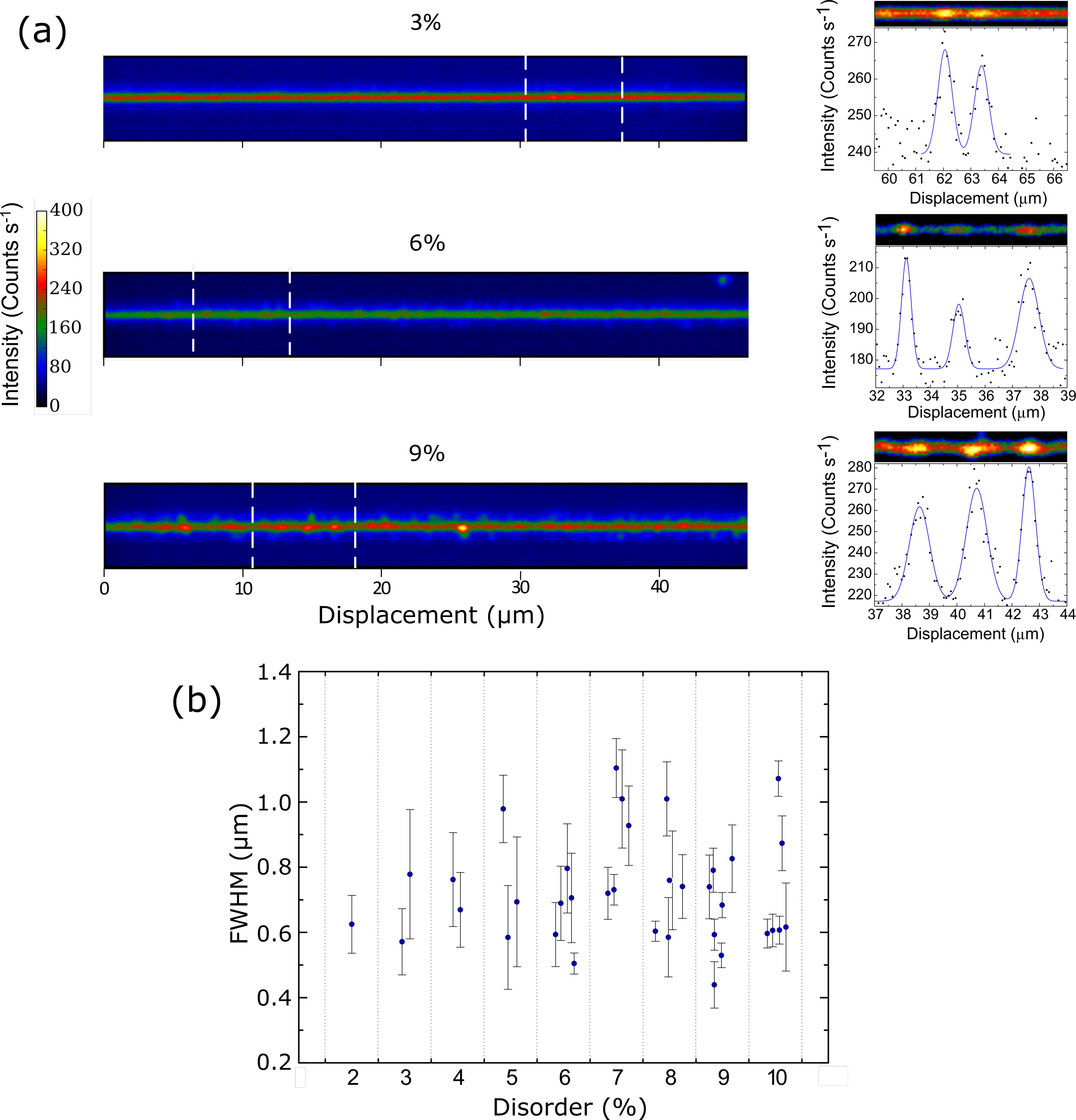}
\caption{(a) Left panels: Photoluminescence images of the emission from waveguides with different degrees of disorder (3\%, 6\%, 9\%), collected at room temperature, under light emitting diode illumination with a power density of 40\,W/cm$^2$. Right panels: Enlargement of the areas highlighted by the dashed lines in the left panels and horizontal linecuts along the centre of the waveguides (circles) and their Gaussian fits (solid lines). (b) Statistics of the spatial extension (Full Width at Half Maximum, FWHM, of the Gaussian fits) of the peaks observed in the photoluminescence images, plotted as a function of the degree of disorder (offset for clarity). The error bars represent the standard deviation of the FWHM extracted from the fits.}
\end{figure}

\begin{figure}[htbp]
\centering
\includegraphics[height=0.9\linewidth, width=0.7\linewidth]{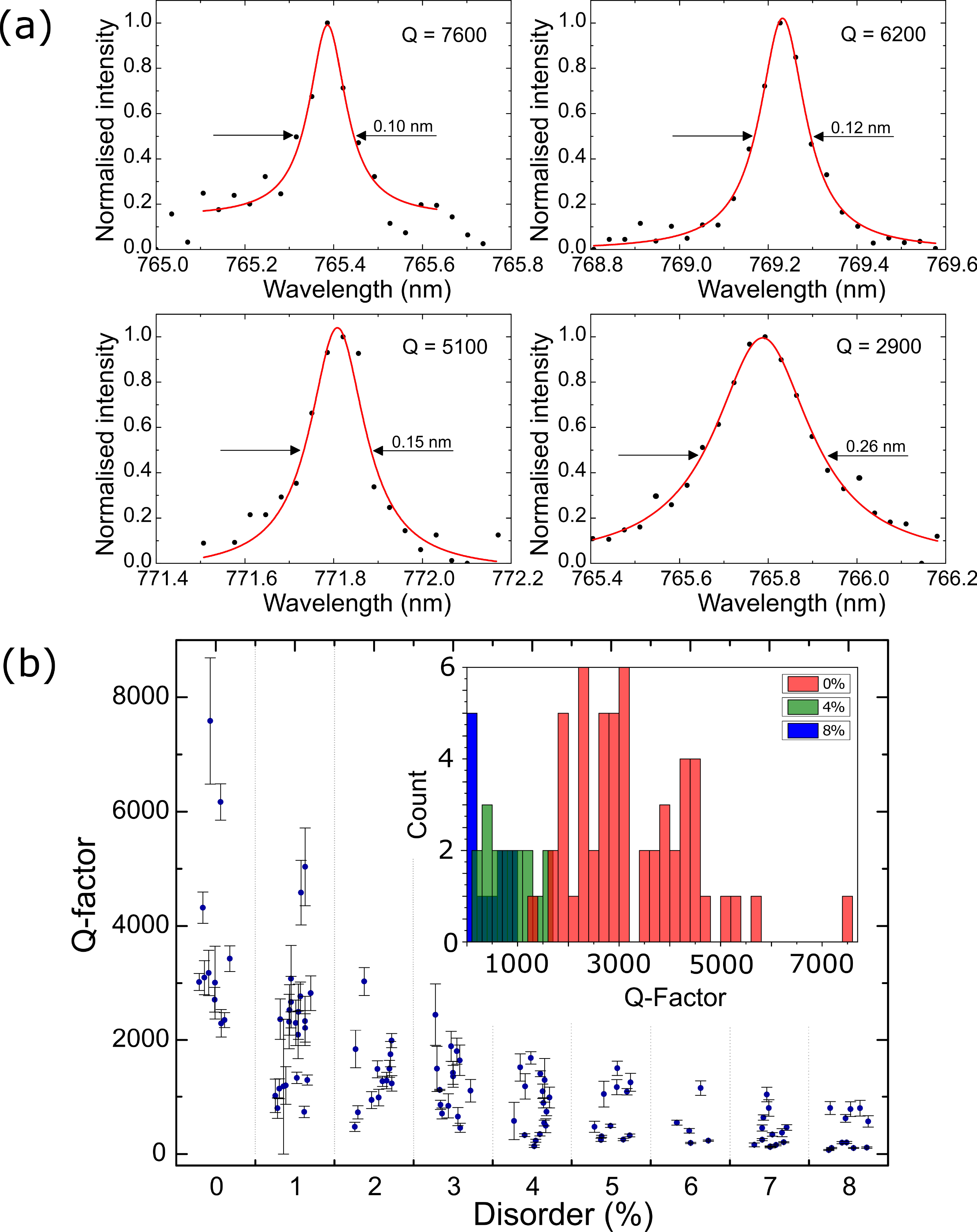}
\caption{(a) Examples of the normalised spectral resonances (symbols) and their Lorentzian fits (solid lines). The arrows indicate the full width at half maxima of the peaks. (b) Statistics of the quality factors as a function of degree of disorder (offset for clarity). The values are extracted from Lorentzian fits, like the ones shown in panel (a), and the error bars represent the standard deviation obtained by propagating the uncertainty in the linewidth and in the central wavelength obtained from the fits. The inset shows histograms of the number of events (count) as a function of quality factor, for different degrees of disorder (0\%, 4\%  and 8\%).}
\end{figure}

\end{document}